\documentstyle[preprint,revtex]{aps}
\begin{document}
\draft
\preprint{January 1995}
\begin{title}
Stability of Insulating Phase in the Chiral Kondo Lattice Model
\end{title}
\author{D. F. Wang and C. Gruber}
\begin{instit}
Institut de Physique Th\'eorique\\
\'Ecole Polytechnique F\'ed\'erale de Lausanne\\
PHB-Ecublens, CH-1015 Lausanne-Switzerland.
\end{instit}
\begin{abstract}
In this work, the stability of the insulating phase
of the 1D chiral Kondo lattice model is studied
at half-filling, within the framework of
self-consistent variational theory.
It is found that arbitrarily small interaction would
drive the system from a conducting phase to an insulating
phase, in spite of the chirality of the conducting band.
\end{abstract}

\narrowtext

Recently, there have been considerable interests in theoretical
studies of Kondo insulators and heavy
fermion systems,
the class of rare earth
compounds\cite{lee,coleman1,coleman2,flude,read,coleman3,yu1,wang,carruzzo,rice,fye,millis,varma,san,aeppli,troyer,tsvelik,ts,white}.
The mechanism for these systems to become insulating
is due to the interaction between the conducting electrons
and the array of localized impurity spins.
On a one dimensional lattice, when the electrons hop between
nearest-neighboring sites, the metal-insulator phase transition
in the system at half-filling described by the Kondo lattice model
has been investigated with
various methods\cite{wang,fye,ts,yu1,tsvelik,troyer}.
It was found that the simple
Fermi liquid fixed point of the system ($J=0$) is unstable
against infinitesimally small interaction. Arbitrarily small
interaction would drive the system from
conducting phase to an insulating phase.

More recently, Carruzzo and Yu develop a novel
self-consistent variational approach, to investigate again
whether an external magnetic field could induce
a metal-insulator phase transition in the Kondo insulators\cite{carruzzo}.
It is found that the insulating gap never vanishes
for any value of magnetic field, suggesting no
field-induced metal-insulator transition,
in contrast to the predictions by the slave boson
mean field theory\cite{millis,varma,san,carruzzo}.

In the self-consistent variational theory,
two scattering processes between the electrons
and the impurity spins are kept in consideration:
the process of  no-spin-flip and zero momentum
transfer ($S_1$), and that of spin-flip and
$\pi$ momentum transfer ($S_2$).
In absence of external magnetic field, the approach
has indicated that the insulating phase is stable
even for arbitrarily small interaction parameter,
that is, $J_c=0^+$, with a small but finite energy
gap decaying exponentially,
consistent with the previous results obtained
through other methods\cite{carruzzo}.
In the following, we use their approach to
study the stability of the insulating phase
of the 1D chiral Kondo lattice model introduced
recently by us. We find that at half-filling, the insulating
phase is stable for any nonzero coupling constant.
The insulating gap is computed explicitly for this system.

The 1D chiral Kondo lattice model consists of conducting
electrons moving in one direction on a close chain.
At each lattice site, the conducting electrons interact
with a localized impurity spin through spin exchange interaction.
The system is described in terms of the following Hamiltonian:
\begin{equation}
H_{ck}=\sum_{k}\sum_{ \sigma=\uparrow,\downarrow}
e(k) c_{k\sigma}^\dagger c_{k\sigma} + J \sum_{i=1}^L
c_{i\alpha}^\dagger {\vec \sigma_{\alpha\beta} \over 2} c_{i\beta}
\cdot {\vec S_f(i)}-h\sum_i[c_{i\alpha}^\dagger c_{i\alpha}
{\sigma_{\alpha\alpha}^z \over 2} + s_f^z(i)],
\end{equation}
where the conduction band spectrum is $e(k)=-tk$, with
$-\pi (1-1/L)\le k\le \pi (1-/L)$,
in the momentum space. $J$ is the coupling
constant between the local impurity moments and the conducting electrons,
and $h$ is the external magnetic field.
The local moments are described by the spin 1/2 operators, that is,
$[S_f^x(i), S_f^y(i)]=iS_f^z(i)$ (plus two other commutation
relations obtained by the cyclic permutations of $x,y,z$),
with the relation $\vec S_f^2(i) =3/4$, for all the sites $i=1, 2, \cdots, L$.
In the following, for simplicity, we always restrict ourselves
to the parameter range $t>0$ and antiferromagnetic
interaction $J\ge 0$. The dispersion relation of the
conducting band $e(k)= -|t| k$ yields the electron group velocity
$v=\partial e(k)/\partial k = -|t|$. Below, we only consider
the thermodynamic limit $L\rightarrow \infty$.

In the extreme limit $J=+\infty$, away from half-filling,
we have found that
the wavefunctions of the system are the RVB-type Jastrow
product wavefunctions\cite{gruber}. The interesting aspect
is that in this limit various correlations of the system
can be computed exactly in compact form. Moreover in this limit,
the system obviously is a conductor away from half-filling, while
it is an insulator at half-filling.
For very large but finite $J$ and at half-filling, each
impurity spin will also attempt to form a singlet with
one conduction electron at each site, to lower the
energy of the system as much as possible.
To transfer one electron with the long range hopping matrix
element (corresponding to the chiral band)
from one site to another would
break two singlets, giving rise to a charge gap of the order
$O(J)$, making the system an insulator.
One interesting issue is the stability of the insulating
phase. In the following, we'll investigate the stability
of the insulating phase of the system when one changes $J$,
and we'll restrict ourselves to the case of half-filling.

Following Popov and Fedotov\cite{popov}, one first introduces
fermion operators $\{f_{j\alpha}^\dagger; f_{j\alpha}\},
j\in\{1, 2, \cdots, L\}, \alpha\in\{+,-\}=\{\uparrow,\downarrow\}$
to describe the impurity spins, together with an additional
term $-{i\pi \over 2\beta} \sum_{j\alpha} f_{j\alpha}^\dagger f_{j\alpha}$
to take into account the constrain that every site is occupied
by one impurity. The partition function of the system thus takes
the following form:
\begin{equation}
Z=\int Dc^\dagger Dc Df^\dagger Df \,\,\,\, e^{-\int_0^\beta d\tau
[\sum_{i=1}^L
\sum_{\alpha=\pm} c_{i\alpha}^\dagger(\tau) \partial_{\tau} c_{i\alpha} (\tau)
+\sum_{i=1}^L \sum_{\alpha=\pm} f_{i\alpha}^\dagger(\tau)\partial_{\tau}
f_{i\alpha}(\tau)+ H(\tau)]},
\end{equation}
where $c$ and $f$ are Grassman variables, and $H(\tau)$
is given in momentum space by
\begin{eqnarray}
H(\tau)&&=\sum_k\sum_{\alpha=\pm} [e(k)-\alpha h/2]c_{\alpha}^\dagger(k,\tau)
c_{\alpha}(k,\tau) + \sum_{k}\sum_{\alpha=\pm}[-{i\pi\over2\beta}-\alpha h/2]
f_{\alpha}^\dagger(k,\tau) f_{\alpha}(k,\tau) +\nonumber\\
&&+{J\over 4L} \sum_{k_1,\cdots,k_4;\alpha,\beta,\gamma,\delta}
\delta (k_1-k_2+k_3-k_4) f_\alpha^\dagger(k_1,\tau) f_\beta(k_2,\tau)
c_\gamma^\dagger(k_3,\tau) c_\delta (k_4,\tau)
\vec \sigma_{\alpha\beta} \cdot \vec \sigma_{\gamma\delta}.\nonumber\\
\end{eqnarray}
Furthermore, introducing ``right" and ``left" operators
for electrons and impurities, i.e.\cite{carruzzo}
\begin{equation}
\phi_{a,\alpha}(k,\tau)=\left\{ \begin{array}{ll}
\phi_{R;\alpha}(k,\tau)=\phi_\alpha(k,\tau)\,\,\,       & \mbox{$k>0,\,\,a=0$
}\\
\phi_{L;\alpha}(k,\tau)=\phi_\alpha (k-\pi,\tau)\,\,\,  & \mbox{$k>0,\,\,a=1$ }
                            \end{array}
\right.
\end{equation}
with $a\in \{R,L\}=\{0,1\}$, and $\phi_{a,\alpha}$ standing either for
$c_{a,\alpha}$ or for $f_{a,\alpha}$, one can
write $H(\tau)$ as follows:
\begin{eqnarray}
H(\tau)=&&\sum_{k>0}\sum_{a=0,1}\sum_{\alpha=\pm} [e(k-a\pi)-\alpha h/2]
c_{a,\alpha}^\dagger(k,\tau)c_{a,\alpha}(k,\tau) +\nonumber\\
&&+\sum_{k>0}\sum_{a=0,1}\sum_{\alpha=\pm} [-{i\pi \over 2\beta} - \alpha h/2]
f_{a,\alpha}^\dagger(k,\tau) f_{a,\alpha}(k,\tau)+\nonumber\\
&&+\sum_{k_1,k_2>0}
\sum_{a,b=0,1}\sum_{\alpha,\beta=\pm} f_{a,\alpha}^\dagger (k_1,\tau)
\tilde C_{\alpha\beta}^{ab} f_{b,\beta}(k_2,\tau),
\end{eqnarray}
where
\begin{eqnarray}
\tilde C_{\alpha\beta}^{ab}=&& {J\over 4L}
\sum_{k_3,k_4>0}\sum_{c,d=0,1}\sum_{\gamma,\delta=\pm}
c_{c,\gamma}^\dagger(k_3,\tau) c_{d,\delta}(k_4,\tau)[\vec
\sigma_{\alpha\beta}\cdot
\vec \sigma_{\gamma\delta}] \times\nonumber\\
&&\times\delta (k_1-k_2+k_3-k_4-(a-b+c-d)\pi),
\end{eqnarray}
with $\vec \sigma_{\alpha\beta} \cdot \vec \sigma_{\gamma\delta}
=\alpha \gamma \delta_{\alpha,\beta} \delta_{\gamma,\delta}
+ 2\delta_{\alpha,-\beta} \delta_{\gamma,-\delta}$,
and we have neglected the $k=0$ terms.

Following Ref.\cite{carruzzo}, we shall keep only the scattering processes:
\begin{eqnarray}
&&S_1:\,\,\,\, k_1=k_2, k_3=k_4, a=b,c=d,\alpha=\beta,\gamma=\delta,\nonumber\\
&&S_2:\,\,\,\, k_1=k_2,k_3=k_4,a=b+1,c=d+1\,\,\,\, (mod 2), \alpha=-\beta,
\gamma=-\delta.
\end{eqnarray}
The two scattering processes correspond to the following effective
interaction:
$S_1: {J\over 4L}
[\sum_{i,\alpha} \alpha f_{i,\alpha}^\dagger f_{i,\alpha}]
[\sum_{j,\gamma}\gamma c_{j,\gamma}^\dagger c_{j,\gamma}],
S_2: {J\over 4L} [\sum_{i,\alpha} (-1)^i f_{i,\alpha}^\dagger
f_{i,-\alpha}][\sum_{j,\gamma} (-1)^j c_{j,\gamma}^\dagger c_{j,-\gamma}].$
Therefore, $\tilde C$ is the $4\times4$ matrix
given in this approximation by
\begin{equation}
\tilde C=\pmatrix{\tilde C_{++}^{ab}&\tilde C_{+-}^{ab}\cr
                  \tilde C_{-+}^{ab}&\tilde C_{--}^{ab}\cr}
=\pmatrix{\tilde C_{++} {\bf 1} & \tilde C_{+-}\sigma_x\cr
          \tilde C_{-+}\sigma_x & -\tilde C_{++} {\bf 1}\cr},
\end{equation}
with matrix elements given by
\begin{eqnarray}
&&\tilde C_{++}={J\over 4L} \sum_{k>0} \sum_{b=0,1}\sum_{\gamma=\pm}
c_{b,\gamma}^\dagger (k,\tau) c_{b,\gamma}(k,\tau)\gamma,\nonumber\\
&&\tilde C_{+-} = {J\over 2L} \sum_{k>0} \sum_{b=\pm}
c_{-b,-}^\dagger(k,\tau) c_{b,+}(k,\tau),\nonumber\\
&&\tilde C_{-+}={J\over 2L} \sum_{k>0} \sum_{b=\pm}
c_{-b,+}^\dagger (k,\tau) c_{b,-}(k,\tau).
\end{eqnarray}
Introducing 4-vector $f^\dagger =(f_{0\uparrow}^\dagger,f_{1\uparrow}^\dagger,
f_{0\downarrow}^\dagger,f_{1\downarrow}^\dagger)$, the interaction term is
thus given by
\begin{equation}
H_I(\tau)=\sum_{k>0} f^\dagger (k,\tau) \tilde C f(k,\tau).
\end{equation}
We can then rewrite the partition function in terms of
following functional integral
\begin{equation}
Z\approx \int Dc^\dagger Dc Df^\dagger Df ~\exp \{-[ S_{oc}[c^\dagger,c]
+S_{of}[f^\dagger,f]+S_{fc}[c^\dagger,c,f^\dagger,f] ]\},
\end{equation}
where the various actions are given by
\begin{eqnarray}
&&\left\{ \begin{array}{ll}S_{oc}={1\over \beta} \sum_{k>0}\sum_{\omega}
c^\dagger(k,\omega) D_0 c(k,\omega)\\
(D_0)_{\alpha\beta}^{ab} =\delta_{a,b}\delta_{\alpha,\beta}
[i\omega -\alpha h/2+ e(k-a\pi)]&\mbox{}
\end{array}
\right.\nonumber\\
&&\left\{  \begin{array}{ll}
S_{of}={1\over \beta} \sum_{k>0}\sum_{\omega} f^\dagger(k,\omega) F_0
f(k,\omega)\\
(F_0)_{\alpha\beta}^{ab} =\delta_{a,b}\delta_{\alpha,\beta}
[ i\omega-{i\pi\over 2\beta } -\alpha h/2 ]&\mbox{}
\end{array}
\right.
\end{eqnarray}
and $\phi(k,\tau)={1\over \beta} \sum_{\omega={(2n+1)\pi\over\beta}}
\phi(k,\omega)e^{i\omega\tau}$.
Keeping only the terms with $\omega_1=\omega_2$ ( thus $\omega_3=\omega_4$ )
as considered in Ref.\cite{carruzzo}, we have
\begin{equation}
\left\{ \begin{array}{ll}
S_{fc} = {1\over\beta^2}
\sum_{k>0} \sum_{\omega} f^\dagger (k,\omega) C f(k,\omega)&\mbox{}\\
C=\pmatrix{C_{++}{\bf 1} & C_{+-} \sigma_x\cr
           C_{-+}\sigma_x&-C_{++}{\bf 1}\cr}, &\mbox{}
\end{array}
\right.
\end{equation}
with matrix elements:
\begin{eqnarray}
&&C_{++}={J\over 4 L \beta} \sum_{k>0} \sum_\omega\sum_{b,\gamma}
c_{b,\gamma}^\dagger (k,\omega) c_{b,\gamma} (k,\omega) \gamma,\nonumber\\
&&C_{+-}={J\over 2L\beta} \sum_{k>0} \sum_\omega\sum_b
c_{-b,-}^\dagger (k,\omega) c_{b,+}(k,\omega),\nonumber\\
&&C_{-+}={J\over 2L\beta} \sum_{k>0}\sum_\omega\sum_b
c_{-b,+}^\dagger (k,\omega) c_{b,-}(k,\omega).
\end{eqnarray}

For this chiral Kondo lattice model, we follow the
self-consistent variational approach of Carruzzo and Yu.
To find the quasi-particle spectrum of the electron
degrees of freedom, one first integrates out the
impurity degrees of freedom, to obtain
the effective action of the electrons. When integrating
out the Popov-Fedotov fermions for the impurity spins,
one replaces the bare propagator with its dressed version,
taking into account the effects of the conduction electrons
on the impurity spins themselves.
After some algebraic
manipulation, we obtain the effective functional integral
for the electron degrees of freedom:
\begin{equation}
Z=\int Dc^\dagger Dc ~ \exp -S_{eff}[c^\dagger, c],
\end{equation}
with the effective action
\begin{equation}
S_{eff} = {1\over \beta} \sum_{\omega} \sum_{k>0}
c^\dagger (k,\omega) D(k,\omega) c(k,\omega),
\end{equation}
where using the same approximation of Ref.\cite{carruzzo},
the corresponding matrix $D$ is found to be
\begin{equation}
D=\pmatrix{ i\omega-\bar h +e(k) &0&0& -{JC_\pi\over 4 \lambda}\cr
            0&i\omega-\bar h +e(k-\pi)& -{JC_\pi\over 4\lambda}&0\cr
            0&-{JC_\pi\over 4 \lambda}&i\omega + \bar h +e(k)&0\cr
            -{JC_\pi\over 4 \lambda}&0&0&i\omega+\bar h +e(k-\pi)\cr},
\end{equation}
where  $e(k)=-kt$, $C_0$ and $C_\pi$ are the mean field parameters
\begin{equation}
C_0=\lim_{\beta\rightarrow\infty} {<C_{++}>\over \beta},
C_\pi=\lim_{\beta\rightarrow\infty} {<C_{+-}>\over \beta},
\end{equation}
and
\begin{equation}
\lambda = [(h/2-C_0)^2+C_\pi^2]^{1/2},\nonumber\\
\bar h =h/2-{J\over 4\lambda} ({h \over 2} -C_0 ).
\end{equation}
Finding the inverse of the matrix $D$, with the analytical
continuation $ i\omega\rightarrow w$, the poles of the propagators
yield the quasiparticle dispersion relations. We have found
that there also exist four bands $E_1(k), E_2(k), E_3(k)$ and
$E_4(k)$, given below:
\begin{eqnarray}
&&E_1(k)=kt-{\pi t/2} - [({\pi t/2} + |\bar h|)^2 +
({JC_\pi/ 4 \lambda})^2]^{1/2},\nonumber\\
&&E_2(k)=kt-{\pi t/2} - [({\pi t/2} - |\bar h|)^2 +
({JC_\pi/ 4 \lambda})^2]^{1/2},\nonumber\\
&&E_3(k)=kt-{\pi t/2} + [({\pi t/2} - |\bar h|)^2 +
({JC_\pi/4\lambda})^2]^{1/2},\nonumber\\
&&E_4(k)=kt-{\pi t/2} + [({\pi t/2} + |\bar h|)^2 +
({JC_\pi/4\lambda})^2]^{1/2}.
\end{eqnarray}

Let us only consider the case of no external magnetic
field $h=0$.
For sufficiently large coupling constant $J$,
when the system is in the insulating phase,
the upper two bands $E_3(k)$ and $E_4(k)$ are positive,
the lower two bands $E_1(k)$ and $E_2(k)$ are negative.
The energy gap of the system is given by
\begin{equation}
\Delta = E_3(0)-E_2(\pi).
\end{equation}
The self-consistent equations for the mean field parameters
$C_0$ and $C_\pi$ are given by:
\begin{eqnarray}
&&\lambda={J^2\over 16} \int_0^\pi {dk\over 2\pi}
[{1\over E_4} + {1\over E_3}]\nonumber\\
&&C_0={J\over 4} \int_0^\pi {dk\over 2\pi}
[{\bar h + \pi t/2 \over E_4 } + {\bar h -\pi t/2 \over E_3}],
\label{eq:self}
\end{eqnarray}
where $\lambda=(C_0^2+C_\pi^2)^{1/2}, \bar h =JC_0/(4\lambda)$.
If the quasi-particle energy gap $\Delta$ is greater
than zero, the insulating phase is stable. However, as the coupling
constant $J$ is small enough, the energy gap approaches to zero,
and the system becomes conducting at this critical point.

In order for the theory to be self-consistent, the definitions
of the mean field parameters $C_\pi$ and $C_0$ must satisfy the condition
$|C_\pi|\le J/4, |C_0|\le J/4$. The self-consistent Eq.~(\ref{eq:self})
can be solved for $C_0$ and $C_\pi$. One can find that
$C_0=0$. The value of $C_\pi$ is determined by the following equation:
\begin{eqnarray}
|C_\pi|&&=(J^2/16) \int_0^\pi {dk\over \pi}
{1\over kt -\pi t/2 +[(\pi t/2)^2 + (J/2)^2]^{1/2}}\nonumber\\
&&={J\over 8} f(|{J\over \pi t}|),
\end{eqnarray}
where the function $f(x)=x\ln{1+[1+x^2]^{1/2} \over x}$.
Since $0\le f(x) <1$ for any $x>0$, one sees that
$|C_\pi|<{J\over 8}$, a physically acceptable solution.
For any coupling $J$, the quasi-particle energy gap at
half-filling is found to be
\begin{equation}
\Delta = -\pi t + [(\pi t)^2 + J^2/4]^{1/2}.
\end{equation}
The above result indicates that the system is always in
in an insulating state for any nonzero coupling $J$ and
$h=0$. Hence, we
conclude that $|J_c|=0^+$, in spite of the chirality of conducting
band ( corresponding to long range hopping matrix ).
In the strong interaction limit, the energy gap is
$\Delta \approx |J|/2$. In the weak interaction limit,
the energy gap becomes $\Delta \approx J^2/(8\pi t)$.

In summary, the stability of the insulating phase
of the 1D chiral Kondo lattice model at half-filling
is studied, within the framework of self-consistent
variational theory. It is found that the insulating phase is
stable for any nonzero impurity-electron interaction,
and the critical point of metal-insulator phase transition
is at $J=0$ in zero external magnetic field.
It would be very interesting to validate the
conclusion of the critical point through other approaches.
Further work is necessary for study of the behavior
of the chiral Kondo lattice model in nonzero magnetic field.

We would like to thank D. Baeriswyl and H. Kunz for conversations.
We are also grateful to H. Carruzzo for correspondences.
This work was supported by the Swiss National
Science Foundation.


\begin{references}
\bibitem{lee} P. A. Lee, T. M. Rice, J. W. Serene,
L. J. Sham and J. W. Willkins, Comments Condens. Matter
Phys. {\bf 12}, 99 (1986).
\bibitem{coleman1} P. Coleman, Phys. Rev. {\bf B30}, 3035 (1984).
\bibitem{coleman2} P. Coleman, Rutgers preprint,
cond-mat/9410024, {\it Questions and issues arising at SCES'94
theory summary-Amsterdam--Strongly correlated electrons}.
\bibitem{flude} P. Flude, J. Keller and G. Zwicknagl,
Solid State Phys. {\bf 41}, 1(1988).
\bibitem{read} N. Read, D. M. News and S. Doniach,
Phys. Rev. {\bf B 30}, 6420 (1984).
\bibitem{coleman3} P. Coleman, E. Miranda and A. Tsvelik,
Phys. Rev. {\bf B 49}, 8955 (1994).
\bibitem{yu1} Clare C. Yu and Steven R. White,
Phys. Rev. Lett. {\bf 71}, 3866 (1993).
\bibitem{wang} Z. Wang, X. P. Li and D. H. Lee,
Phys. Rev. {\bf B47}, 11935 (1993).
\bibitem{carruzzo} H. Carruzzo and Clare C. Yu,
UC Irvin preprint (1994), cond-mat/9412050.
\bibitem{rice} T. M. Rice and K. Ueda, Phys. Rev. {\bf B34}, 6420 (1986).
\bibitem{fye} R. M. Fye and D. J. Scalapino,
Phys. Rev. {\bf B44}, 7486 (1991).
\bibitem{millis} A. J. Millis, in {\it Physical Phenomena
at High Magnetic Fields,}, eds. E. Manoussaki et al.
(Addison Wesley, Redwood City, 1992) P. 146.
\bibitem{varma} C. M. Varma, Phys. Rev. {\bf B 50}, 9952 (1994).
\bibitem{san} C. Sanchez-Castro et al, Phys. Rev. {\bf B47}, 6879 (1993).
\bibitem{aeppli} G. Aeppli and Z. Fisk, Comments Cond.
Mat. Phys. {\bf 16}, 155 (1992).
\bibitem{troyer}
M. Troyer and D. Wurtz, Phys. Rev. {\bf B47}, 2886 (1993).
\bibitem{tsvelik} A. M. Tsvelik, Oxford-preprint, cond-mat/9405024.
\bibitem{ts} H. Tsunetsugu, Y. Hatsugai, K. Ueda and M. Sigrist,
Phys. Rev. {\bf B 46}, 3175 (1992).
\bibitem{white} S. R. White, Phys. Rev. Lett. {\bf 69}, 2863 (1992).
\bibitem{popov} V. N. Popov and S. A. Fedotov,
Sov. Phys. JEPT {\bf 67}, 535 (1988).
\bibitem{gruber} D. F. Wang and C. Gruber,
IPT-EPFL preprint, submitted to Journal of Statistical Physics.
\end{references}
\end{document}